\documentclass[sigconf]{acmart}

\pdfoutput=1

\usepackage{cleveref}

\DeclareFixedFont{\ttb}{T1}{txtt}{bx}{n}{10} 
\DeclareFixedFont{\ttm}{T1}{txtt}{m}{n}{10}  

\usepackage{color}
\definecolor{deepblue}{rgb}{0,0,0.5}
\definecolor{deepred}{rgb}{0.6,0,0}
\definecolor{deepgreen}{rgb}{0,0.5,0}

\usepackage{subfigure}
\usepackage{graphicx}

\usepackage{listings}

\newcommand\pythonstyle{\lstset{
language=Python,
basicstyle=\ttm,
morekeywords={self},              
keywordstyle=\ttb\color{deepblue},
emph={MyClass,__init__},          
emphstyle=\ttb\color{deepred},    
stringstyle=\color{deepgreen},
frame=tb,                         
showstringspaces=false
}}
\newcommand\pythoninline[1]{{\pythonstyle\lstinline!#1!}}

\AtBeginDocument{%
  \providecommand\BibTeX{{%
    \normalfont B\kern-0.5em{\scshape i\kern-0.25em b}\kern-0.8em\TeX}}}

\acmConference[ICSE-SEIP 2022]{Submitted to ICSE '22: Software Engineering in Practice}{June 03--05, 2018}{Pittsburgh, PE}
\acmBooktitle{Woodstock '18: ACM Symposium on Neural Gaze Detection,
  June 03--05, 2018, Woodstock, NY}
\acmPrice{15.00}
\acmISBN{978-1-4503-XXXX-X/18/06}

\begin{document}

\newcommand{\kaboom}{Kab{\sc oom}} 
\newcommand{\tsne}{t-{\sc sne}}
\newcommand{\oom}{{\sc oom}}
\newcommand{\ios}{i{\sc os}}

\newcommand{\todo}[1]{
    \fcolorbox{black}{yellow}{\bfseries\sffamily\scriptsize todo}
    {\sf\small$\blacktriangleright$\emph{#1}}
  }

\title{KabOOM: Unsupervised Crash Categorization through Timeseries Fingerprinting}

\author{Edward Yao}
\affiliation{%
  \institution{Facebook}
    \city{Menlo Park}
  \state{CA}
  \country{USA}
}
\email{edwardcdy@fb.com}

\author{Wes Dyer}
\affiliation{%
  \institution{Facebook}
    \city{Menlo Park}
  \state{CA}
  \country{USA}
}
\email{wesdyer@fb.com}

\author{Georgios Gousios}
\affiliation{%
  \institution{Facebook}
  \city{Menlo Park}
  \state{CA}
  \country{USA}
}
\email{gousiosg@fb.com}

\renewcommand{\shortauthors}{Yao et al.}

\begin{abstract}
Modern mobile applications include instrumentation
that sample internal application metrics at regular intervals.
Following a crash, sample metrics are collected 
and can potentially be valuable for root-causing difficult to diagnose crashes. 
However, the fine-grained nature and overwhelming wealth of available application metrics, 
coupled with frequent application updates,
renders their use for root-causing crashes extremely difficult.

We propose \kaboom{}, a method to automatically cluster
telemetry reports in intuitive, distinct crash categories.
Uniquely, \kaboom{} relies on multivariate timeseries fingerprinting;
an auto-encoder coupled with a cluster centroid optimization technique
learns embeddings of each crash report,
which are then used to cluster metric timeseries based crash reports.
We demonstrate the effectiveness of \kaboom{} on both reducing the 
dimensionality of the incoming crash reports
and producing crash categories that are intuitive to developers.

\end{abstract}



\maketitle

\section{Introduction}

The reliability of modern applications, 
running either on the cloud or on mobile devises, 
is paramount for their success.
Despite heavy investment in software quality processes,
including testing, static analysis, and code reviews,
bugs are still propagated to production-level systems, hampering user experience.
Bugs may manifest as application crashes,
whose triaging, root causing and fixing 
demands strong expertise on how the application is structured and how it works.
As such crashes happen in environments not fully controlled by the application developers,
debugging engineers often can only rely on telemetry to root cause the crashes.

The problem we focus on is the debugging of Facebook iOS application crashes that happen as a result of the application running out of memory (OOM). An OOM crash happens when the Facebook mobile application consumes memory above a certain specified threshold allocated by the operating system, which then causes the operating system to kill the application. These issues, while usually relatively infrequent, can occasionally affect a significant portion of the user base as result of buggy code or configuration changes. In some cases, a code change causes a new type of OOM crash, which we term an emerging crash. Particularly, when these crashes occur in popular portions of the application, engineers wish to debug these crashes as quickly as possible; every hour saved during the debugging process can result in significantly less user crashes. 

Compared to other types of crashes, OOM crashes do not usually produce actionable stack traces, which would normally help developers localize and quickly debug the crash. Consequently, OOM crashes are some of the hardest to debug. To alleviate the lack of signal, the Facebook suite of mobile applications contain telemetry modules that provide useful information collected from applications before they crash. Telemetry works by sampling internal application object metrics at regular intervals, or at developer specified points in the application lifetime. When a crash is detected and the user gives permission, those samples are uploaded to Facebook's internal systems for further analysis. Debug engineers can then stitch together individual samples from each mobile application to form a \emph{multivariate timeseries} per session, representing its object allocation count and overall memory leading up to the time of the crash. 

Currently, engineers tasked with debugging OOM crashes manually comb through dozens to hundreds of crashing sessions, for each session examining a visual plot of the object metric allocation counts over time. This data is hard to comprehend, as each session can have hundreds of object allocation timeseries with distinct behavior. 
With luck, engineers are able to rely on previous knowledge and heuristics to notice correlations between when an application hits the maximum memory limit and anomalous behavior in one or a few object allocations across many different sessions. 
In turn, such correlations may give hints as to the root cause of the OOM error.
However, combing through hundreds of crashes containing hundreds of allocation
timeseries in order to spot such patters is taxing for engineers, 
which is why OOM crashes can take long to root cause.




In this work, we present \kaboom{}, a method and a corresponding implementation to automatically cluster
application crashes using timeseries fingeprinting. The inputs to the \kaboom{} model are multivariate time series, where each univariate timeseries is the count of a particular object
allocations
over a crashing session's lifetime. \kaboom{} uses an autoencoder model to embed the time series into a clustering space, where similar crash embeddings are closer to each other and unrelated crash embeddings are far away from each other.
For training,
\kaboom{} assumes that a relatively small (currently, under 20) number of
different crash types exist in the input space.
To ensure maximum cluster separability, 
it conditions its encoder on crash types found on early samples immediately
after deployment for a new version.
In production, the first output of \kaboom{} is the cluster into which an input session belongs to.
The second output is a list of likely "important" object metrics per cluster, which is obtained by using our model's cluster label assignments to run object-level comparisons between the clusters.

We apply the \kaboom{} model on a real-world use case, 
specifically \ios{} {\sc oom} crashes from the Facebook application over the course of five weeks.
Since \kaboom{} is an unsupervised model for a novel application,
our evaluation is constrained by the lack of both ground truth data
and competing solutions.
We thus proceed to demonstrate the effectiveness of \kaboom{}
through a series of analyses of its ability to capture important aspects of the problem at hand.
Initially, we compare the clustering ability of different embedding model configurations, 
without and with the clustering module.
Then, we evaluate \kaboom{}'s ability to learn to identify emerging types of crashes,
i.e., crash types that are new between successive application versions or
successive times.
Finally, we demonstrate how \kaboom{} can help engineers in root causing 
\oom{} crashes.

This work makes the following concrete contributions: 

\begin{itemize}
    
    \item The concept of \emph{timeseries fingerprinting} for 
    analyzing crashing application sessions.
    
    \item \kaboom{}, an end-to-end unsupervised pipeline that uses
    timeseries fingerprinting to cluster crashes.
    
    \item A method for explaining assignments of incoming application
    traces to particular clusters.
    
\end{itemize}

\begin{figure*}[t]
    \centering
    \includegraphics[width=1\textwidth]{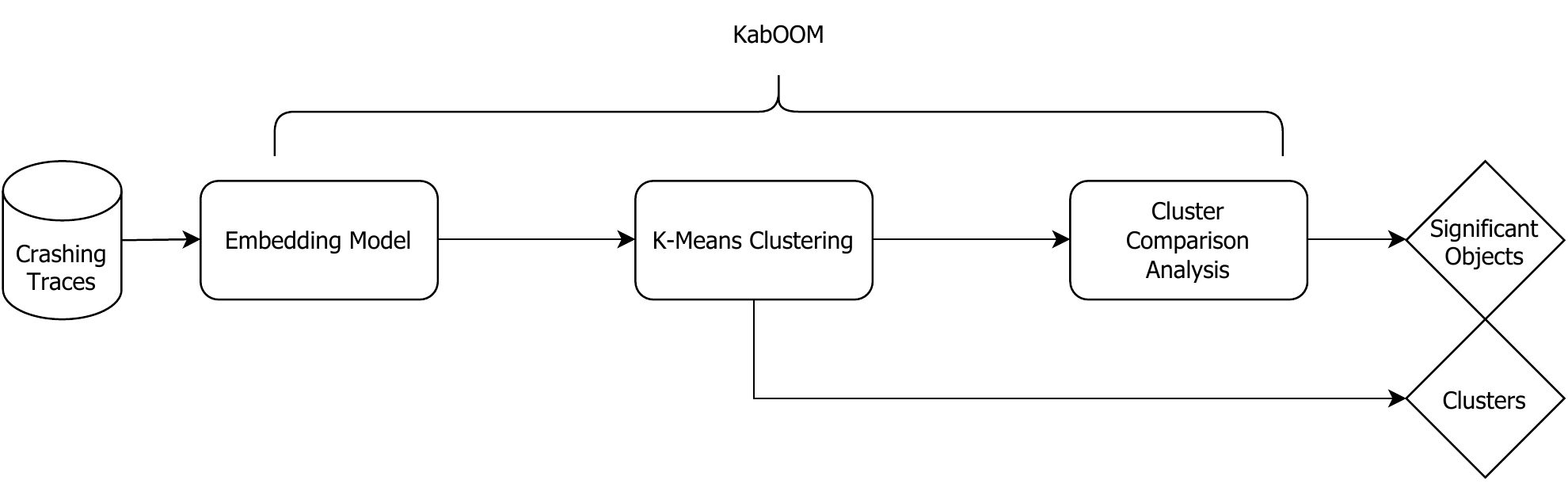}
    \caption{System Diagram \textmd{KabOOM takes in crashing traces. The embedding model (the encoder portion of an autoencoder) is the part of the model that is trained, and learns to embed OOM traces. \kaboom{} then runs K-Means clustering on embeddings of a validation dataset, which assigns each crash to one of $k$ (user-chosen) clusters. Finally, \kaboom{} runs cluster contrastive comparison between the $k$ generated cluster labels and outputs important memory objects per cluster.}}
    \label{fig:overall}
\end{figure*}

\section{Clustering Crash Sessions}

As applications evolve, so do their failures. 
New application versions introduce new paths to failure
and it is precisely those new paths that are interesting for engineers that debug crashes.
It is important from a quality perspective
that new crashes which have a significant impact on application stability or
are experienced by many users to be caught as early as possible.
We therefore model crash categorization as an unsupervised learning problem,
as we do not know a priori the crash types or their number.

In a nutshell, our approach,
\kaboom{}, clusters similar crashing application sessions based on object allocation timeseries
and, per cluster, surfaces those allocation timeseries that led to the crash represented by a cluster.

\kaboom{} works in three phases:

\begin{itemize}
    \item  In the \emph{training} phase, it learns a model to embed (fingerprint) multivariate timeseries. 
    Assuming a single platform (e.g., \ios{}), and to account for changes in application behaviours due to updates, 
    \kaboom{} models need to be trained and deployed on each new application version.

    \item In the \emph{calibration} phase, \kaboom{} uses the trained models to process a sufficiently
    large sample of current traces in order to instantiate representative crash clusters
    (triggered nightly).

    \item In the \emph{production} phase, it continuously processes incoming traces and assigns
    them to the pre-instantiated clusters.

\end{itemize}

\subsection{Input data}
\label{sec:inputdata}
An incoming application trace is an $t \times m$ matrix, 
where $t$ represents the sampling time and
$m$ represents the available metrics. 
Depending on how long an application had been running
prior to the crash, 
$m$ and $t$ can be in the order of 100s or 1000s of metrics and samples, respectively.
It is important to note that as traces are a result of user activity,
not all traces have the same metrics.
We therefore introduce a homogenization step that removes all timeseries that are not
present in more than a configurable threshold of traces.
In case a timeseries is missing from a trace,
we add it as an zero-filled vector. 
Moreover, we trim all timeseries to a configurable number of timesteps,
starting from the most recent measurement; 
we assume that the root cause of a crash is probably manifested
a few timesteps before the crash happened.
If an object is present but for fewer than $m$ timesteps, we zero-pad
the readings from the left of the matrix, i.e. the oldest readings.
Finally, for each input trace, all object timeseries are individually scaled 
with feature range of $[0,1]$.

\begin{figure*}
\subfigure[Stacking Autoencoder]
{
    \includegraphics[scale=0.45]{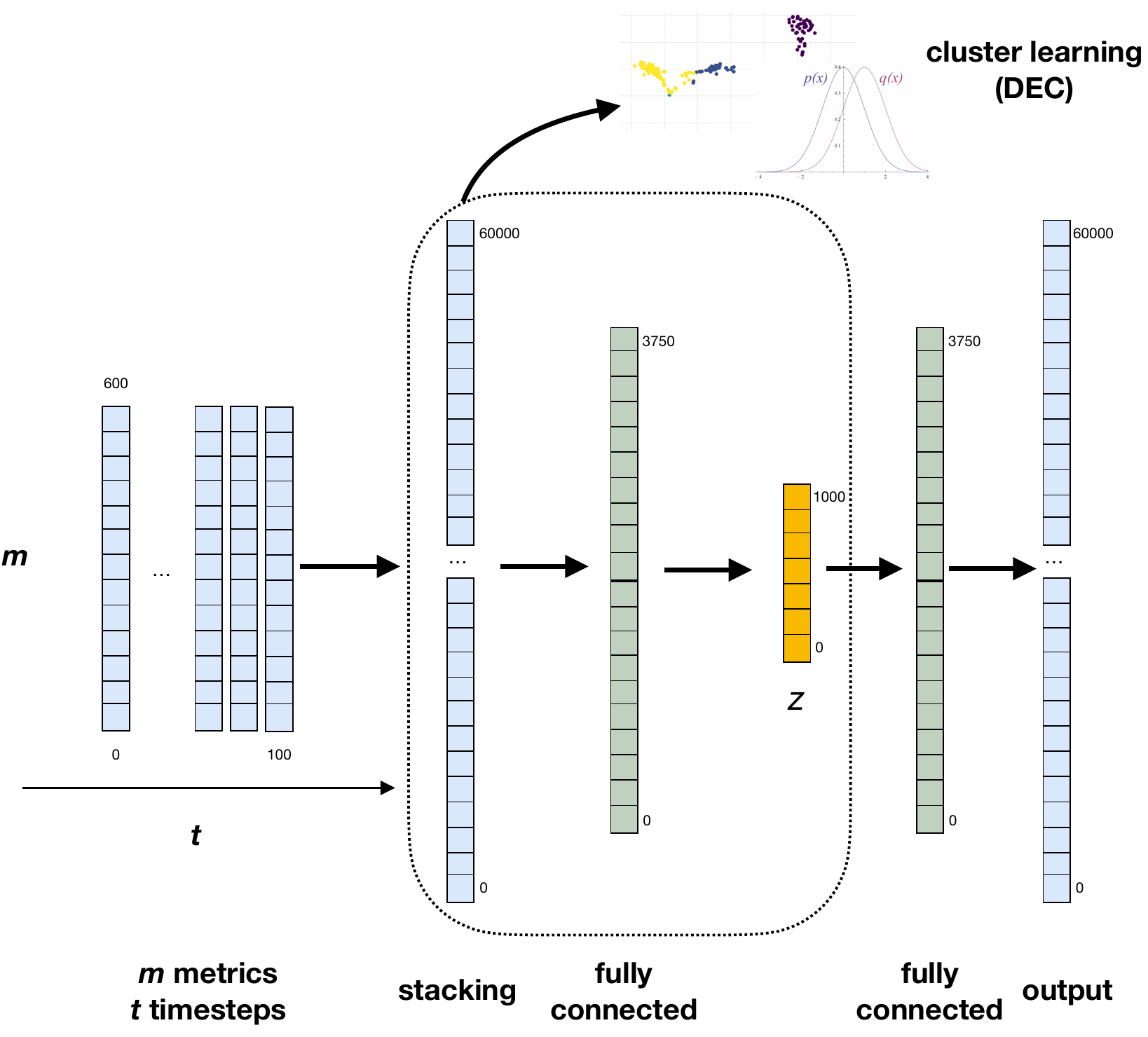}
}
\unskip\ \vrule\ 
\subfigure[Stacking Variational Autoencoder]
{
    \includegraphics[scale=0.45]{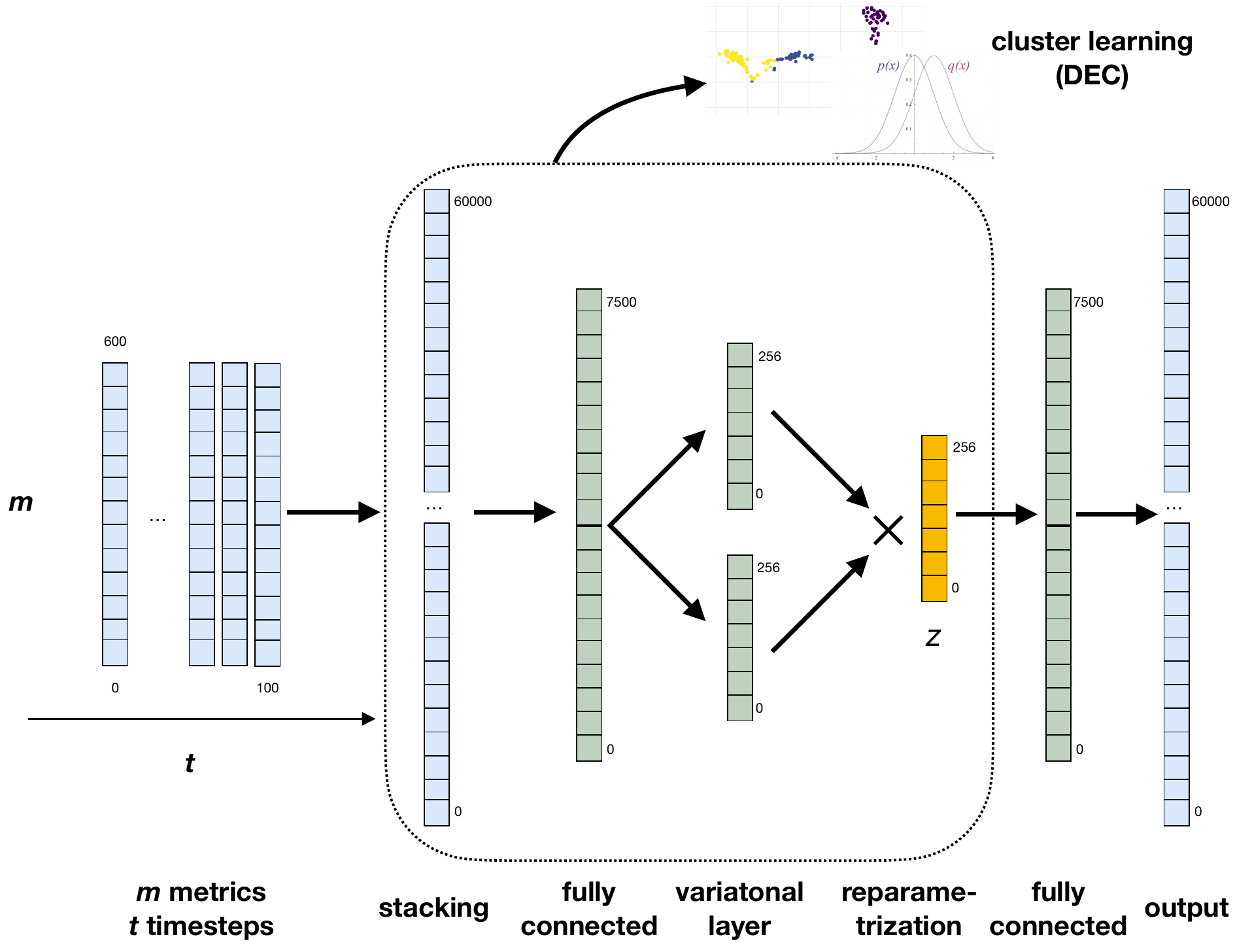}
}
\caption{Architecture of the \kaboom{} autoencoders. 
Lengths are indicative, but proportionally correct. 
In both cases, the encoder part is further conditioned by a cluster learning module.}
\label{fig:model}
\end{figure*}


\subsection{Embedding Model - Training}
Clustering on high dimensional data is impractical. 
Initially, the data is sparse, which makes identification
of appropriate clusters difficult.
The goal of the embedding step is therefore to reduce the input dimensionality to a very compact representation,
while maintaining enough information for good cluster separation.
To do so, the input trace is fed into an autoencoder.

Depending on how the data is fed to the model, 
we can identify two types of models:

\begin{itemize}

    \item Stacking models: Stacking models stack all $m$ dimension
     vectors (columns) on top of each other. The input to the model is an 
     one-dimensional vector, whose length is $n \times m$. Effectively,
     stacking models process the whole trace in one step.

    \item Window models: Window models slide a window of length $s$ along 
    the $t$ dimension, resulting in $t - s$ inputs of size $s \times m$.
    Window models need multiple steps to process a trace, but can process
    longer traces.

\end{itemize}

\begin{table}[t]
\caption{Typical sizes in the \kaboom{} setting}
\label{tab:sizes}
\begin{tabular}{llr}
\toprule
\multicolumn{3}{l}{\emph{Input data}} \\
    & Timesteps                           &  1,000 -- 10,000 \\
    & Metrics                             &  2,500 -- 3,500 \\    
    & Timesteps (after reduction -- $t$)  &  100 \\
    & Metrics (after reduction -- $m$)    &  600 -- 700 \\
    & \# training samples                 &  5,000 \\
\midrule
\multicolumn{3}{l}{\emph{Stacking Model}} \\
    & Input length                    &  $t \times m$ \\
    & AE $z$ length                   & $\dfrac{t \times m}{64}$ \\
    & VAE $z$ length                  & 256 \\
\bottomrule
\end{tabular}
\end{table}


After experimentation with both types of models,
we decided to exclusively use stacking models,
as they are both faster to train and lead to better
cluster separation for our main task.
Our description therefore focuses on stacking models.

\paragraph{Basic Autoencoder}
Autoencoders (AEs) are a class of unsupervised representation learning models developed in deep learning. Generally, autoencoder models learn compact representations of supplied input examples by reconstructing them in a much smaller dimensional space compared to the input space.

Autoencoders comprise two separate networks:
the encoder network takes inputs $x$ from some original input space and 
maps them to a (usually lower-dimensional) latent space specified by the researcher, to produce $z = f_{\theta_f}(x)$. 
The decoder network then takes $z$ as input and maps that back the original input space, 
to produce $\hat{x} = g_{\theta_g}(z)$, where $\hat{x}$ is hopefully similar to $x$. Choices for the structure of the networks represented by $f$ and $g$ vary, with the simplest choices being traditional feed-forward neural networks, and more complex networks employing recurrent and convolutional layers, as well as any other techniques seen across deep learning models.

Both the encoder and decoder networks are trained using stochastic gradient descent to minimize some notion of distance between $\hat{x}$ and $x$, commonly the mean squared error (MSE) 
between each input $x$ and its reconstructed output $r$, where MSE is calculated as~\cite{bengio2013representation, goodfellow2016deep}: 
\[
    MSE=\frac{1}{n}\sum_{i=1}^n(x_i-\hat{x_i})^2
\]

\kaboom{}'s stacking AE is a straightforward fully-connected
AE. It comprises 2 fully connected layers (see Figure \ref{fig:model}). 
The input and the output layer have $t \times m$ dimensions.
The input dimensions are first progressively reduced until 
the embedding layer $z$ and then expanded.

\paragraph{Variational Autoencoder}
In the \kaboom{} setting, simple stacking AEs face a specific problem:
the input dimensionality is too high ($\sim$70,000 dimensions)
to effectively compress with 1 - 2 intermediate layers. 
As the data is very sparse, 
more layers give the AE the opportunity to overfit and
make the model too big to fit on a GPU with reasonable VRAM.
Also, as the $z$ dimension is a function of the input size,
the resulting embeddings are still too long for efficient clustering at scale.
For those reasons, 
we also introduced VAEs as an alternative representation learning technique.

VAEs, proposed by Kingma et. al.~\cite{kingma2014auto}, 
combine variational Bayesian inference methods with the autoencoder architecture
to better capture stochasticity within input data. 
As opposed to learning a static mapping from the input space to latent space, 
VAEs learns a probabilistic distribution over the training data.
Underlying VAEs is the assumption that the input data are distributed using a prior known distribution;
for continuous data, the Gaussian distribution is often assumed by default.
The encoder portion of a VAE, often denoted $q(\cdot)$ in the literature, approximates $q_\theta(z|x) \approx P(z | x)$.The decoder portion of a VAE, often denoted $p(\cdot)$ in the literature, approximates 
$p_\phi(x|z) \approx P(x | z)$. In this notation, $\theta$ and $\phi$ parameterize the encoder and decoder respectively, and can be thought of as the neural network weights and biases which training optimizes over.
$P(x | z)$ is usually assumed to also take on a normal form, and is sampled from a normal distribution parameterized by $z$.
VAEs are trained by minimizing the negative evidence lower bound (ELBO). 
When the prior distribution is assumed to be Gaussian, the ELBO can be calculated as follows:
\[
    ELBO(\phi, \theta) = - \mathbb{E}[\text{log } p_\phi(x|z)] + D_{\text{KL}}(q_\theta(z) || P(z))
\]

Intuitively, the first term represents how likely the observed data is given our choice of latent variable $z$,
and the second term is there to penalize our model if its choice for the latent distribution ($q_\theta(z)$)
strays too far from our assumed functional prior ($P(z)$), where distance here is measured by Kullback–Leibler 
divergence between the two distributions. Just like traditional autoencoders, stochastic gradient descent is used to minimize ELBO loss while training a VAE.

\kaboom{}'s VAE is shallow: 
the input layer downscales the input by a factor of 8 
and then feeds two vectors which, when combined, 
learn the parameters for a Gaussian approximation of the input distribution.
We chose a Gaussian prior to simplify the calculation of the ELBO loss,
using the reparametrization trick as described in 
~\cite{kingma2014auto}.
The variational nature of the VAE model enables us to compress the input 
to a high degree, keeping similar inputs close in the embedding space;
in turn, clustering on those should produce better cluster separation.



\subsection{Clustering Model - Training}
In addition to the embedding models described above, 
\kaboom{} uses an additional model to fine tune the encoders,
so that the generated embeddings are more amenable to clustering.

The way that \kaboom{} improves the embeddings is based on the 
Deep Embedding Clustering (DEC) model~\cite{Xie2016DEC}.
DEC expects an encoder model and an initial set of cluster
centroids (which also implies that the number of clusters $k$ must be fixed before training the model). DEC then proceeds as follows. 

For every iteration, DEC calculates the distance (which we call $Q_{ij}$) of each embedded point $z_i$ (crashing session in our case) to each of the provided cluster centroids $\mu_j$, using a $t$-distribution kernel. This quantity represents how well the model-produced embeddings fit with the currently chosen cluster centroids. Then, DEC calculates a chosen "target distribution," (which we call $P_{ij}$) which encodes how close we want each point $z_i$ to be to a cluster centroid $\mu_j$, and represents what we want embeddings to look like. The choice of the distribution of $P$ is crucial and intuitively should accomplish the following two things: have each point be close to a centroid and have clusters be relatively similar in size.

 DEC's objective is to minimize the 
 Kullback–Leibler divergence between 
 the distributions $Q_{ij}$ and $P_{ij}$
 (i.e. $\sum_i\sum_jKL(P||Q)$). Back propagating the KL loss updates to both the cluster centroids and the encoder concludes one iteration of the algorithm.
Effectively, DEC simultaneously conditions the encoder to produce embeddings that are easier to cluster, and moves the centroids of each cluster to produce better clusters. The cluster centroids calculated by DEC can then be used to initialize K-Means.

The K-Means algorithm, and hence the DEC model, requires the number of clusters $k$ to which to assign
incoming data points. 
A heuristic method to select a $k$ appropriate for a given dataset
is the so-called ``elbow'' method:
KMeans is run for a range of $k$s, 
from which we select the one at which 
a metric (usually the Shiloutte coefficient, see~\Cref{sec:metrics})
stops improving at a significant rate.
In our case, we cannot apply the elbow method on the non-embedded datapoints
as this would be computationally expensive,
but we still need to provide a $k$ for calculating the initial 
cluster centroids we provide as input to DEC.
To solve this problem, we set this initial $k$ 
to a high number (currently 20), 
so we effectively train DEC to learn at least $k$ different cluster centroids.

\subsection{Initializing Clusters - Calibration}

During the calibration phase, after the neural networks are done from both the autoencoder/variational autoencoder reconstruction loss and the DEC clustering loss, we fix the neural networks. We then initialize clusters which we can compare production requests (i.e. live crash data) against. 

To determine the number of clusters our model will be using, we employ the elbow method - we run the K-Means clustering algorithm on the embeddings of held-out validation dataset generated using our AE model and compare clustering metrics across different choices of cluster numbers to find an optimal $k'$ for the particular deployment (see Figure \ref{fig:elbow}). Note that usually $k > k'$, where $k$ is fixed at usually 20 during neural network training time as mentioned previously.

\subsection{Explaining cluster assignments - Production} 
There were not straightforward ways to provide explanations for our model's decisions made in determining cluster assignments, a problem commonly seen across many deep learning approaches. We investigated providing interpretations for the clusters by treating our VAE model as a black box and applying methods accordingly.

The simplest way to we examined the resulting clusters was to run some pattern mining techniques across the data from different clusters. We did this by taking each cluster in turn and comparing the presence of certain features in that cluster versus all other clusters combined. We examined three types of features for all the clusters: the presence of objects (i.e. if the objects had any non-zero readings in a cluster), the lack of presence of objects, and the average values of objects across the clusters.


\subsection{Deployment - Production}

A new model is trained and deployed for each new application version, to account for 
internal changes in the application code.
The deployment of a \kaboom{} model for a particular version is delayed 
until enough data for a particular version has been collected.
After deployment for a particular app version, 
\kaboom{} accepts a crashing session and 
outputs a label indicating the cluster the crash belongs in.
Those labels are integrated in a crash analysis tool that
helps engineers query crashes according to those labels.

\section{Evaluation}

The problem \kaboom{} is trying to solve lays in the field of unsupervised
learning. 
We cannot know neither whether the clusters that \kaboom{} proposes 
are optimal for a given dataset,
nor whether placing a particular crash in a particular cluster is correct.
Therefore, to evaluate \kaboom{}, we cannot rely on existing labeled data, 
as those do not exist.
Moreover, the focus of our work is not to develop a novel multivariate
timeseries clustering method,
but to provide a working solution for our debugging engineers.
We therefore only evaluate \kaboom{} with variations of itself and
refrain from performing a full-blown evaluation with competing 
timeseries clustering 
models.\footnote{Note that our literature research did not reveal any method for clustering multivariate timeseries.}
Our evaluation strategy is therefore mostly qualitative in nature.

Our evaluation is guided by the following research questions:

\begin{itemize}
    \item \textbf{RQ1} Which model best separates crash clusters?
    \item \textbf{RQ2} Can \kaboom{} identify emerging types of crashes?
    \item \textbf{RQ3} How can \kaboom{} explain crash clusters?
\end{itemize}

Our evaluation is performed on a machine with 56 vCPUs, 256GB RAM and
2 Nvidia Tesla P100-SXM2 GPUs with 16GB VRAM.

\subsection{Dataset} Our dataset includes \oom{} error related crashes 
from the Facebook \ios{} application collected over a period of 3 weeks.
The crashes correspond to a single major application version.
From this period we sample over 2,500 
crashes.\footnote{We cannot reveal the exact number of crashes due to company policy}
Each crashing session contains object allocation counts at different points 
in time, with their size ranging in the order of several hundred timeseries.
The original data is split into two sets: a training set ($\approx$2,500 crashes), 
and a hold-out validation set ($\approx$200 crashes). 
We use the training set to train the embedding and clustering models,
and the validation set for evaluation.


Within a major version, the Facebook \ios{} application receives frequent updates, 
which may introduce new types of crashes.
During one of the three weeks from which we sampled the training data, 
a memory related bug caused a set of crashes, which was quickly fixed. 
We denote crashes related to this bug as emerging regression traces. 
Since the \kaboom{} deployment servicing the release 
has not been trained to detect this particular crash, 
it should have difficulties assigning those crashes to its existing clusters.
Consequently, emerging regression traces help us test whether our model 
can flag emerging types of crashes.

Finally, we collect crashing sessions from a period of 2 weeks right after the train set (and subsequently the regression traces), which we call "later crashes". These sessions serve as a comparison point against which we can assess how well our model differentiates regression data versus "normal" looking data, and whether there are natural drifts in the time series data over time or whether data is relatively stationary.

All sessions are trimmed to 100 timesteps and 592 object timeseries, 
using the process outlined in \Cref{sec:inputdata}.

\subsection{Metrics}
\label{sec:metrics}
As we are tackling an unsupervised task with no ground truth labels, 
we resort to similarity and participation based metrics. 
Specifically, for each clustering assignment, we compute the following metrics:

\begin{description}
    
    \item[Silhouette Coefficient] measures the mean distance of a single data point to
    all other points in the cluster and compares it to the mean distance of the same point
    to all items of the closest cluster~\cite{rousseeuw1987silhouettes}. 
    Ranges from -1 (incorrect clustering) to 1 (very dense clustering).
    
    \item[Calinski-Harabasz Index] measures the ratio of the sum of between-clusters 
    dispersion and of within-cluster dispersion~\cite{calinski1974dendrite}. 
    A higher value indicates well-separated clusters.
    
    \item[Davies-Bouldin Index] is based on the ratio between within-cluster and between-cluster distances~\cite{davies1979cluster}. 
    If two clusters are close together, but have a large spread, 
    then this ratio will be large, indicating that these clusters are
    not very distinct. A low value indicates a better clustering.
    The Davies-Bouldin metric can be used in addition to the Silhouette Coefficient
    when performing elbow analysis.
    
\end{description}

It should be noted that all metrics above are known not to perform well in certain
scenarios involving particular data distributions or clustering algorithms~\cite{halkidi2001clustering}.
Consequently, their interpretation is tied to the dataset under evaluation,
so they cannot be used to compare approaches across datasets.

We accompany the analysis of all metrics above with visual inspection of \tsne{} plots. 
\tsne{} is a dimensionality reduction technique~\cite{van2008visualizing} 
whose core property is strong separation of unrelated data points in a (usually) two
dimensional space.

\subsection{Method} 

To measure which of the \kaboom{} model variants 
creates the best cluster assignments (\textbf{RQ1}), we train them using default settings
on our training set. Specifically, we test 4 variants: i) AE, ii) VAE, iii)
 AE + DEC, and iv) VAE + DEC.
We fix the number of epochs to 1000 to ensure a reasonable training time (less than 40 minutes
in all cases, using a single GPU) and set the embedding size $z$ to 256 for VAE models
and 925 for AE ones.
We train each of the four proposed models in turn and then embed the held-out validation dataset using the trained models. We then directly compare cluster metrics and the \tsne{}
visualizations.

To determine whether \kaboom{} can identify emerging types of crashes (\textbf{RQ2}), 
we select the best performing model including the optimal number of clusters 
and we use it to embed the emerging crashes dataset. 
We compare the results to the data in the validation set,
using the same \tsne{} plot.
If there is a noticeable difference in the visualization,
i.e., \tsne{} plots the emerging crashes away from existing crash clusters,
we can conclude that \kaboom{} embeddings are useful for distinguishing 
``normal'' crashes (which engineers already know of) 
from new ``abnormal'' crashes (which engineers should investigate).

Finally, we examine the output of running various tools on the output of the clusters, 
as detailed in section \ref{explainability} (\textbf{RQ3}). 
It is rather hard to determine whether cluster explanations generated by tools make sense, 
even for those with significant domain knowledge 
about \tsne{}, \oom{}s and the Facebook \ios{} application code. 
We use some prior knowledge about the Facebook \ios{} application code 
to assess how reasonable the generated explanations are.

\section{Results}

\subsection{RQ1: Cluster representation learning}
\begin{figure}
    \centering
    \includegraphics[width=0.48\textwidth]{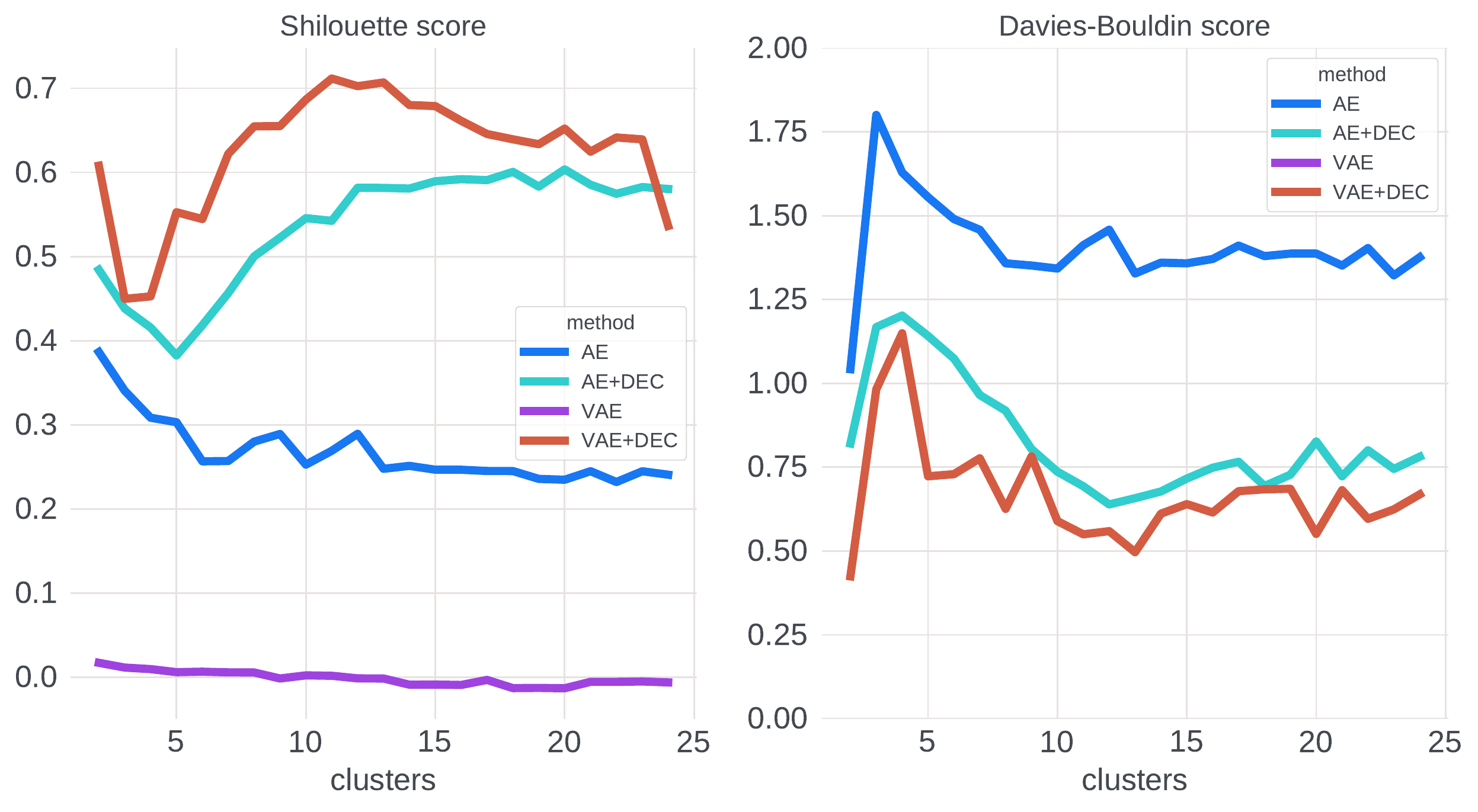}
    \caption{Elbow Analysis. \textmd{Plot of silhouette score and Davies-Bouldin Index for different numbers of clusters supplied to K-Means for the iOS OOM dataset.
    }}
    \label{fig:elbow}
\end{figure}

Before performing clustering with an algorithm like K-Means, 
we need to determine the optimal number of clusters through elbow analysis.
The results for our \ios{} dataset can be seen in~\Cref{fig:elbow}. 
We can immediately observe that the two variations of the models
that were trained with the cluster learning addon (DEC) have significantly
higher Shilouette score and significantly lower Davies-Bouldin score.
This indicates that our strategy to condition our embedding generation for
clustering has proved worthwhile.
It is interesting to observe that before the application of DEC,
the VAE method would have to be discarded based on its terrible scores
(we omitted the VAE line from the Davies-Bouldin plot for clarity --- 
it was consistently over 6);
after applying DEC though, the VAE+DEC model is significantly better
that all other variations.
Both metrics are optimized across the board at about 10-12 clusters.

\begin{figure*}
\subfigure[AE (5 clusters)]
{
    \includegraphics[width=.32\textwidth]{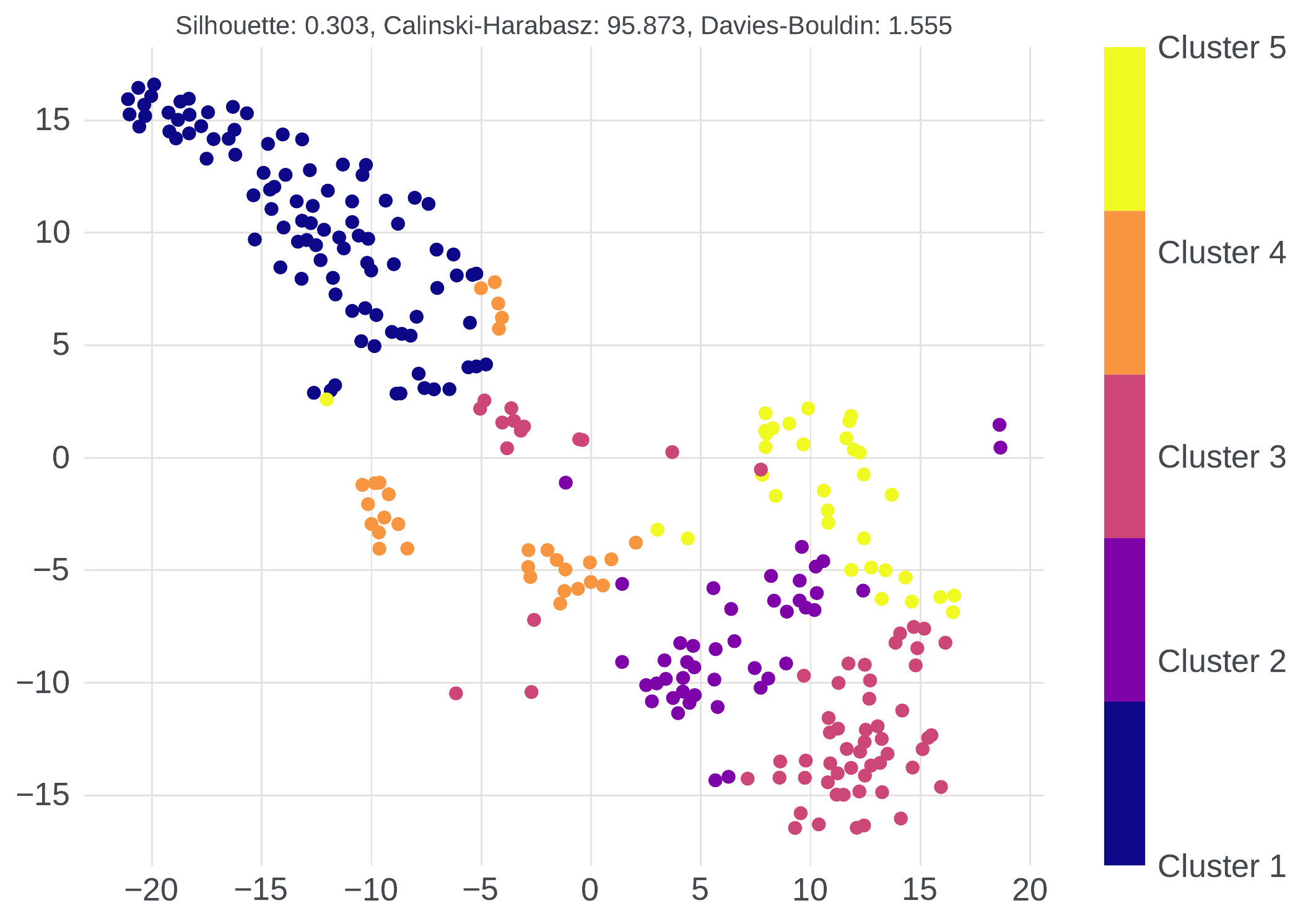}
}
\subfigure[AE + DEC (5 clusters)]
{
    \includegraphics[width=.32\textwidth]{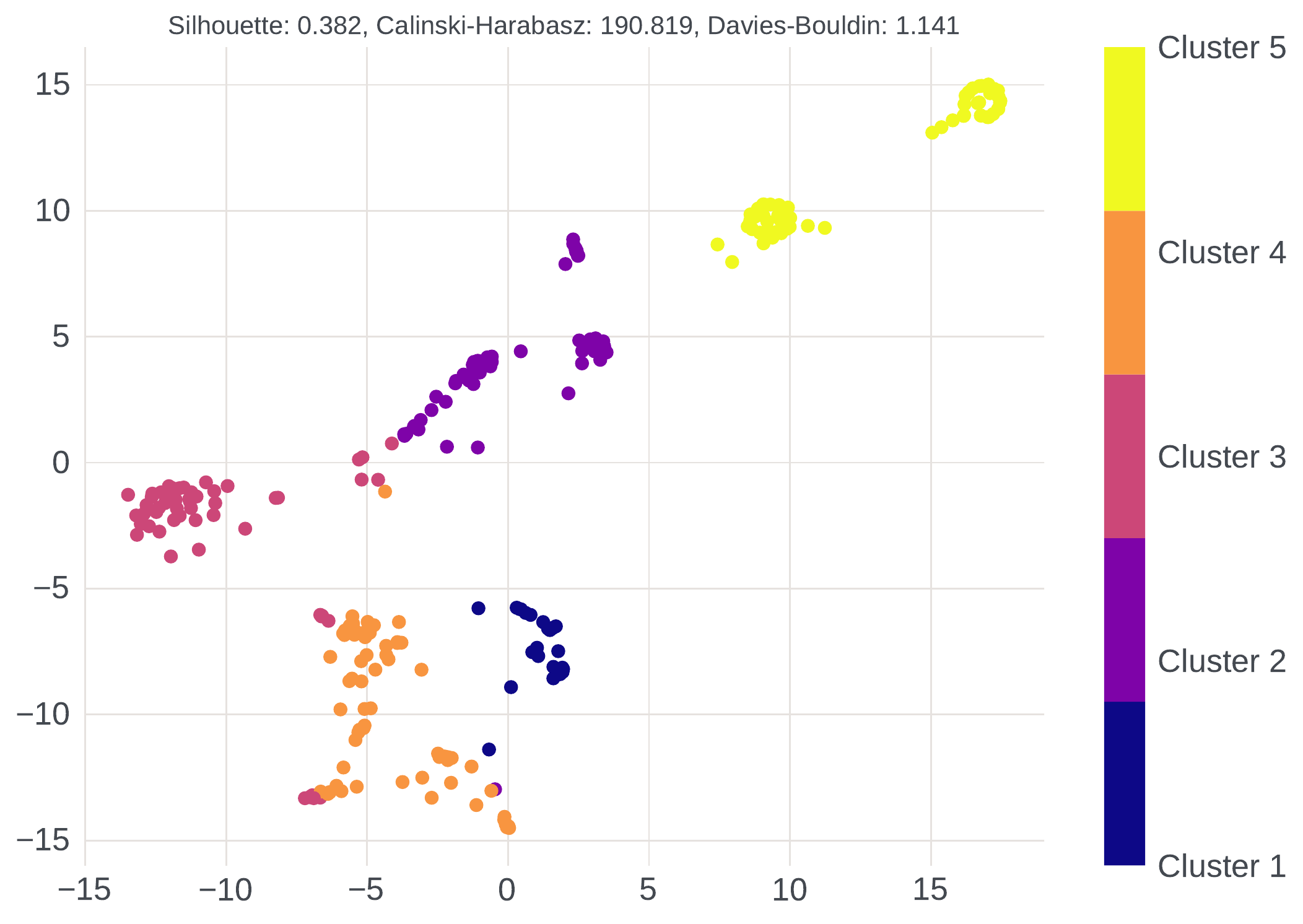}
}
\subfigure[AE + DEC (11 clusters)]
{
    \includegraphics[width=.32\textwidth]{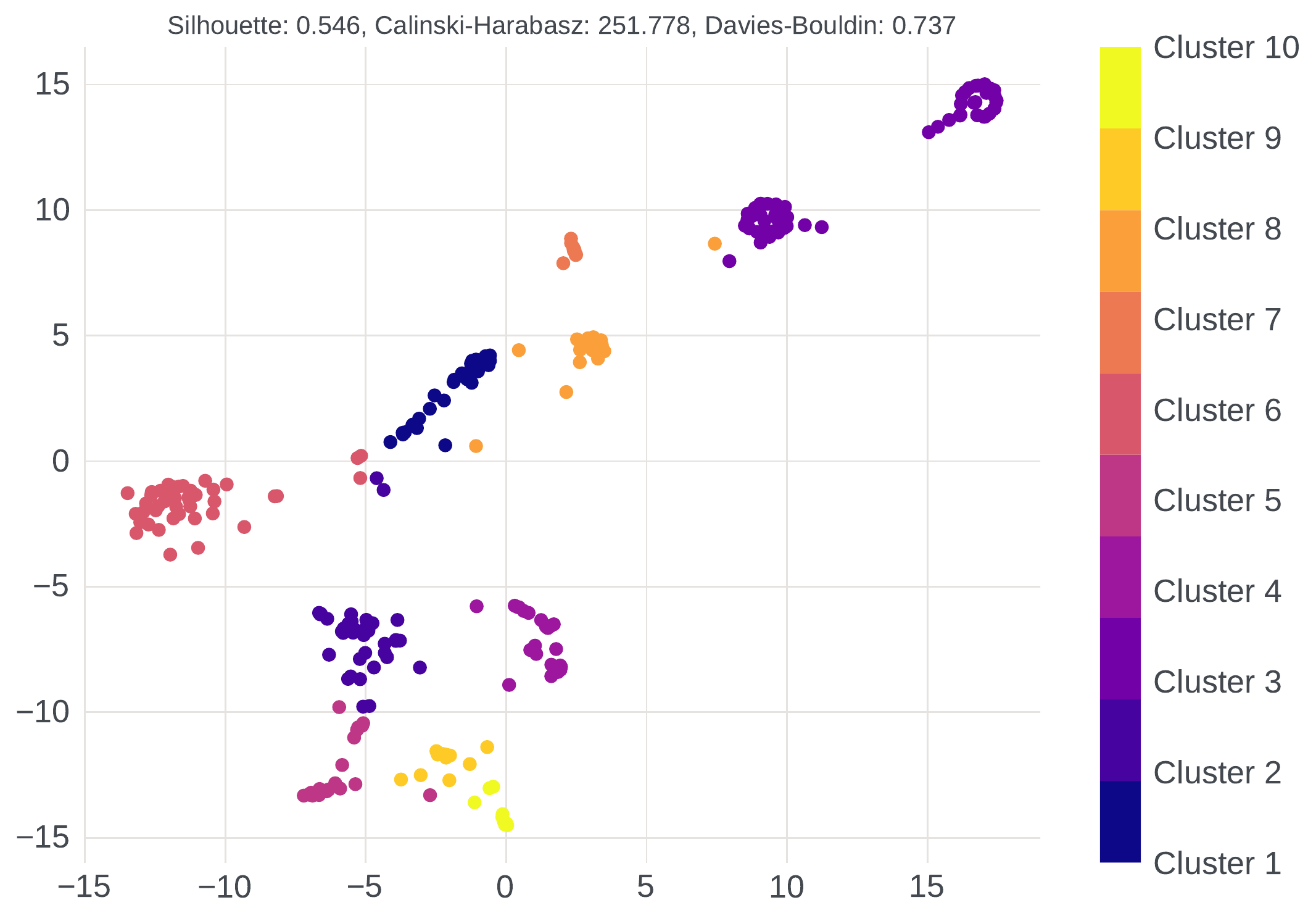}
}
\subfigure[VAE]
{
    \includegraphics[width=.32\textwidth]{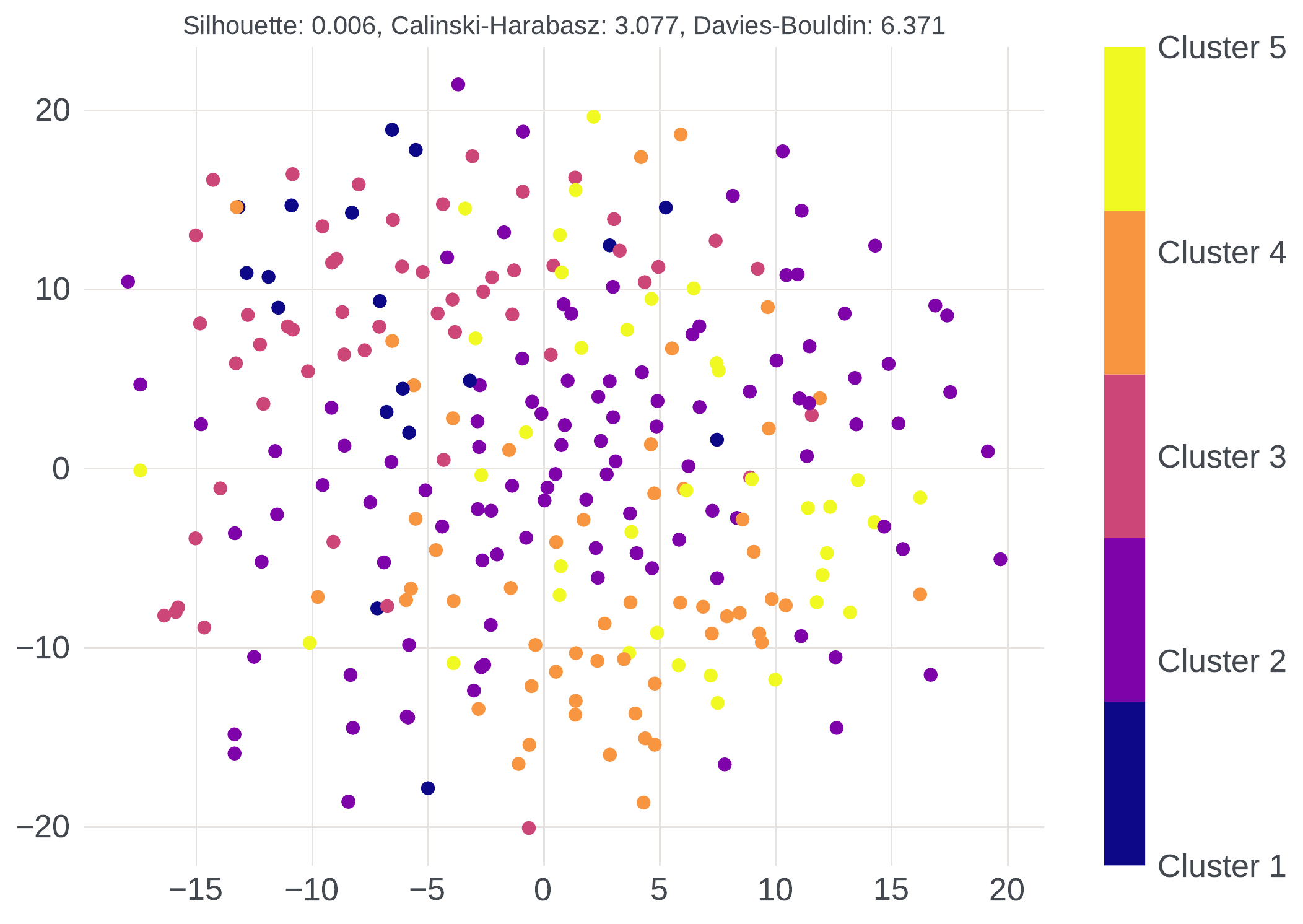}
}
\subfigure[VAE + DEC (5 clusters)]
{
    \includegraphics[width=.32\textwidth]{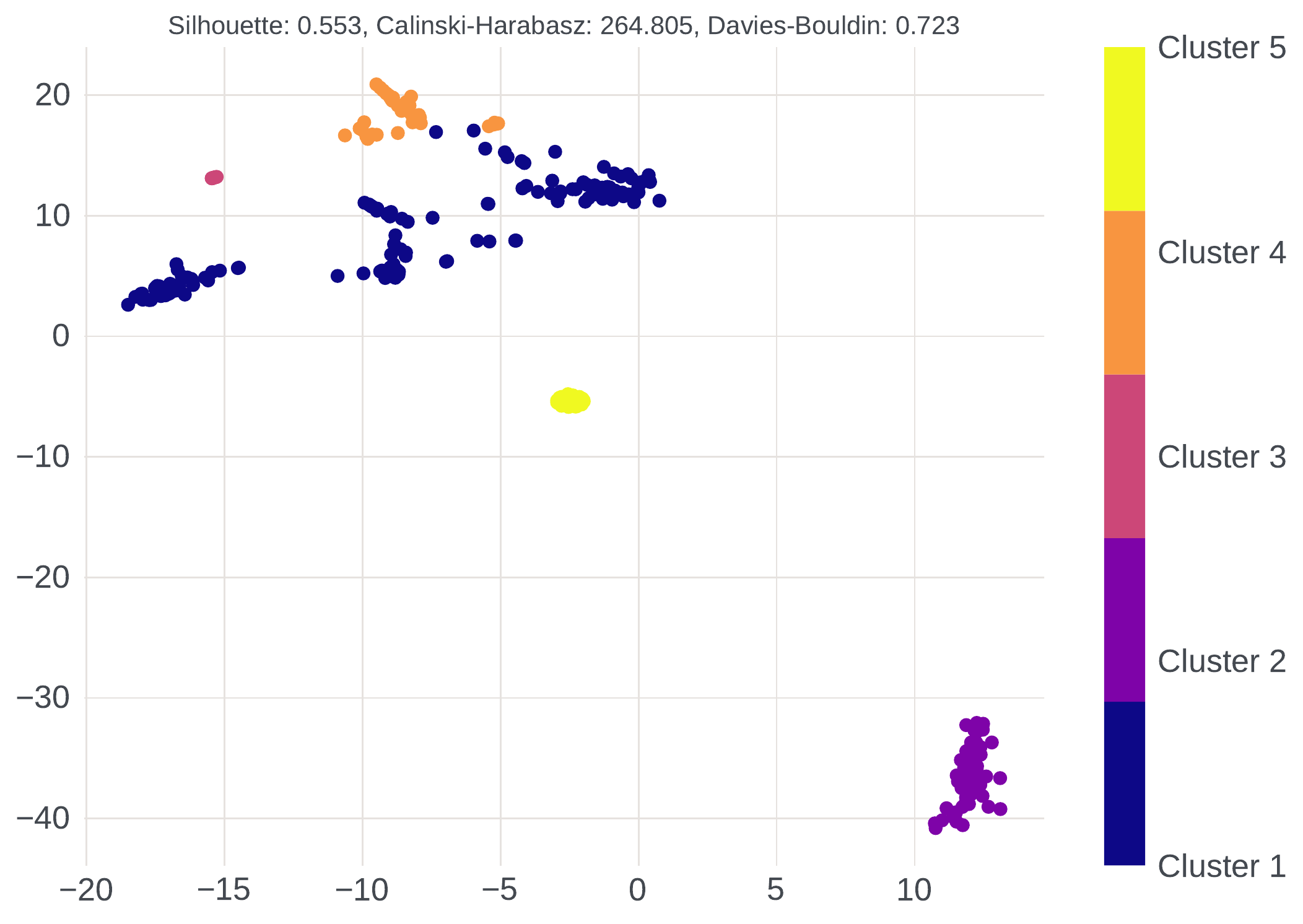}
}
\subfigure[VAE + DEC (11 clusters)]
{
    \includegraphics[width=.32\textwidth]{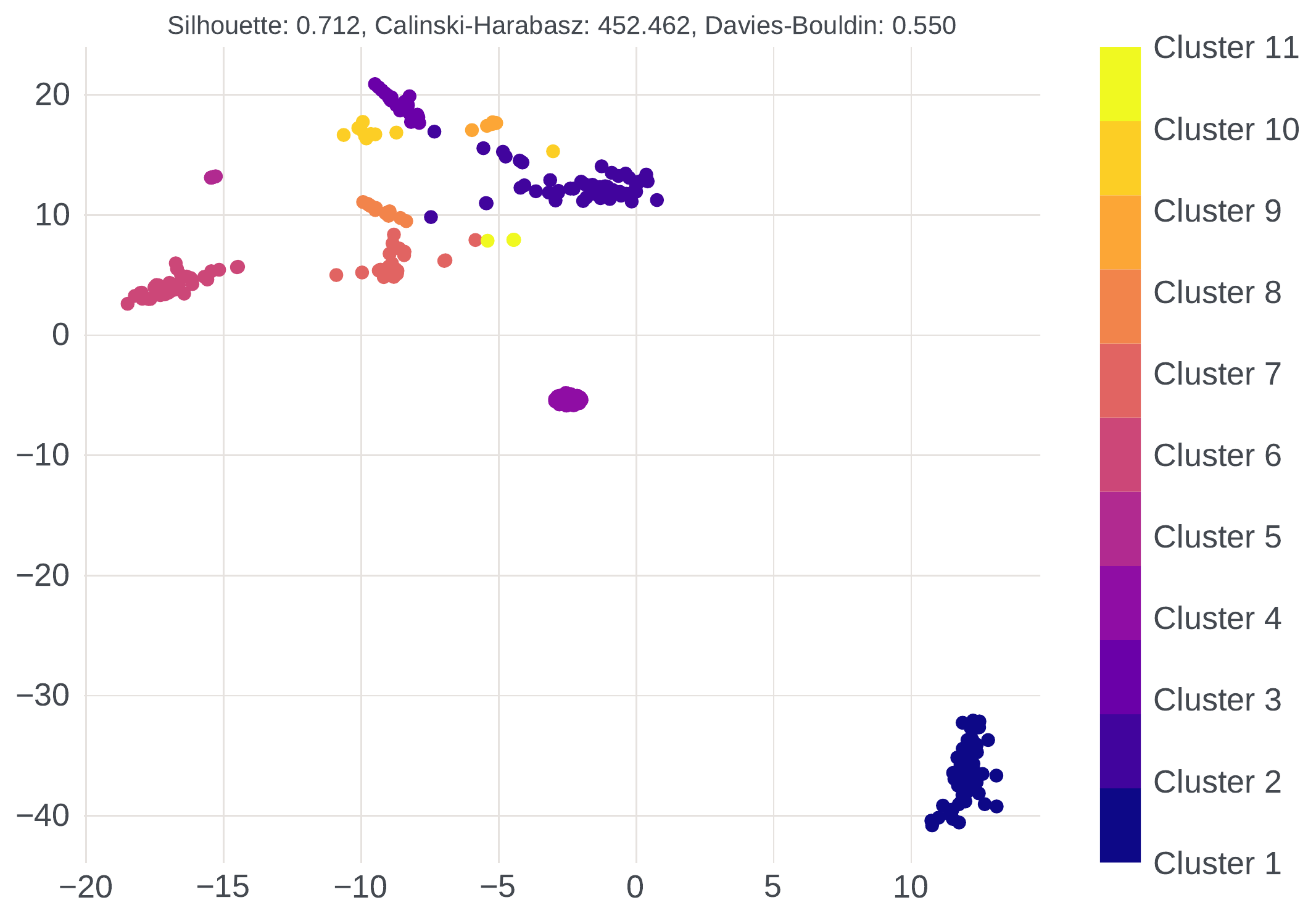}
}

\caption{Cluster separation for various versions of the \kaboom{} autoencoder.}
\label{fig:RQ1}

\end{figure*}

For the second part of RQ1, 
we visually compare the embeddings generated by each clustering method (see~\Cref{fig:RQ1}).
A good embedding should enable KMeans to put similar crashes together,
which would be depicted on a \tsne{} plot as a single colored ``bubble'' with no overlaps.
Without DEC, the AE embeddings seem to work better than the VAE ones;
we see in~\Cref{fig:RQ1}(a) that the AE embeddings form relatively consistent
clusters, whereas the VAE ones, in~\Cref{fig:RQ1}(b), are not. 
The application of DEC makes the clusters more tight in both cases.
The corresponding scores (as indicated at the top of the Figures) are
significantly in favour of the VAE+DEC solution, especially 
in ths case of 11 clusters. 
This is despite the fact that the VAE
embeddings have 75\% less dimensions (256 vs 975), 
which is desirable for computing those clusterings effectively.
This could be due to the fact that the DEC loss function conditions 
the parameters in the VAE's sampling distribution to produce
embeddings much closer to the learned cluster centroids.

It should be noted here that those results may be different
in other datasets (e.g., other application versions), 
but in practice we have observed that the VAE+DEC solution
performs the best for our use case.

\subsection{RQ2: Identifying emerging crash types} \label{RQ2}

\begin{figure*}
\subfigure[Initial clusters]
{
    \includegraphics[width=.32\textwidth]{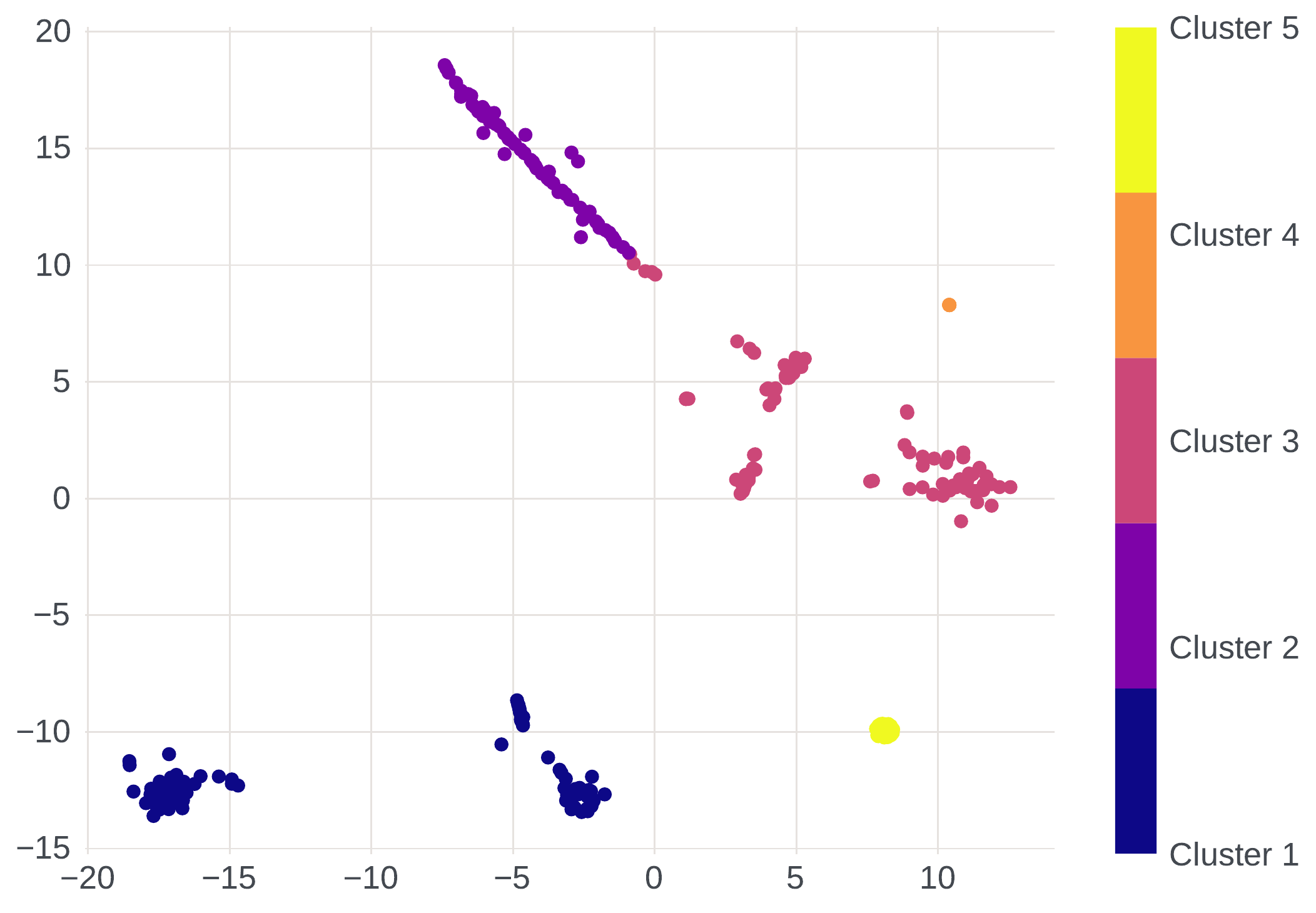}
}
\subfigure[Initial clusters and an emerging regression]
{
    \includegraphics[width=.32\textwidth]{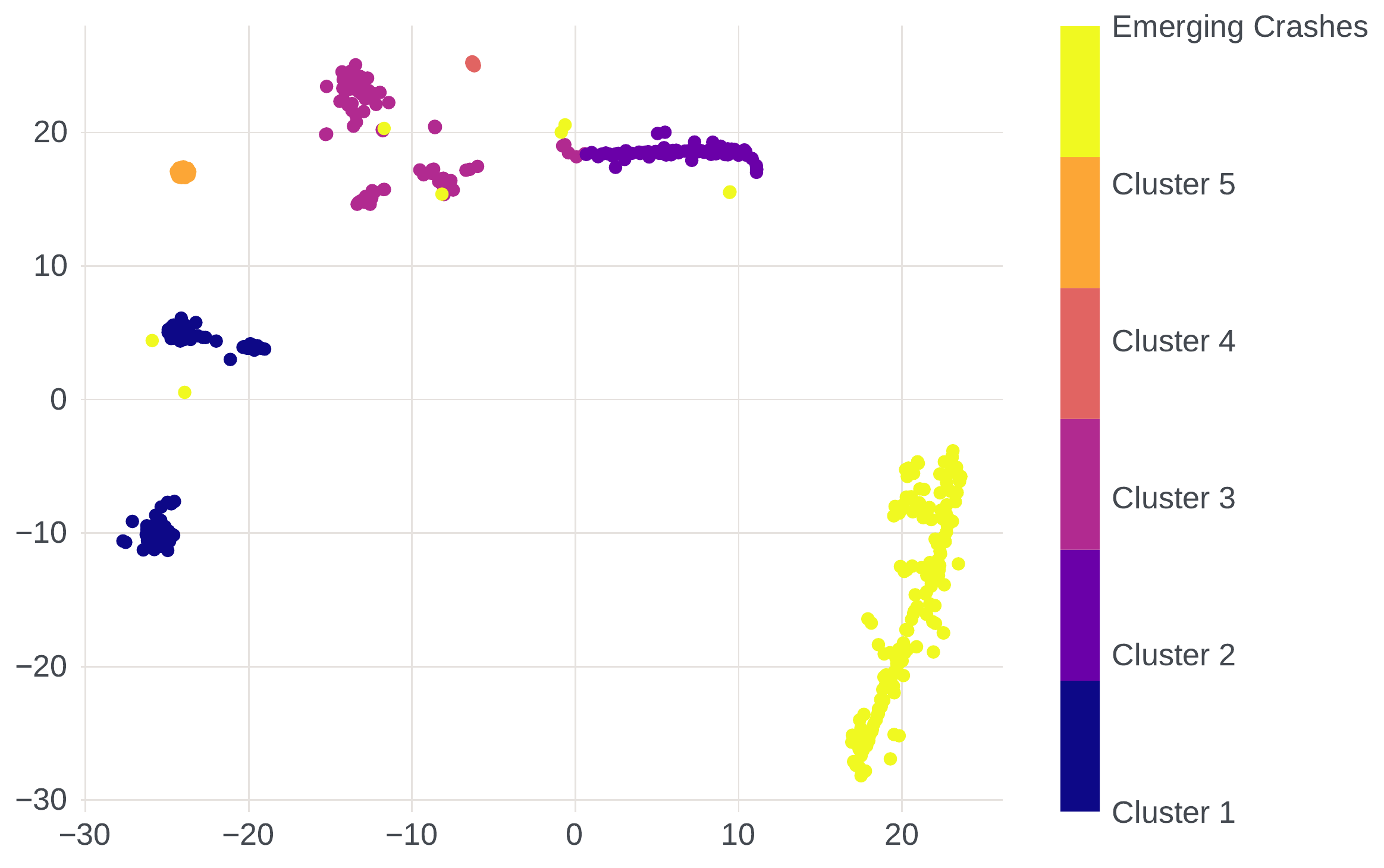}
}
\subfigure[Initial clusters, the previous regression and newer crashes]
{
    \includegraphics[width=.32\textwidth]{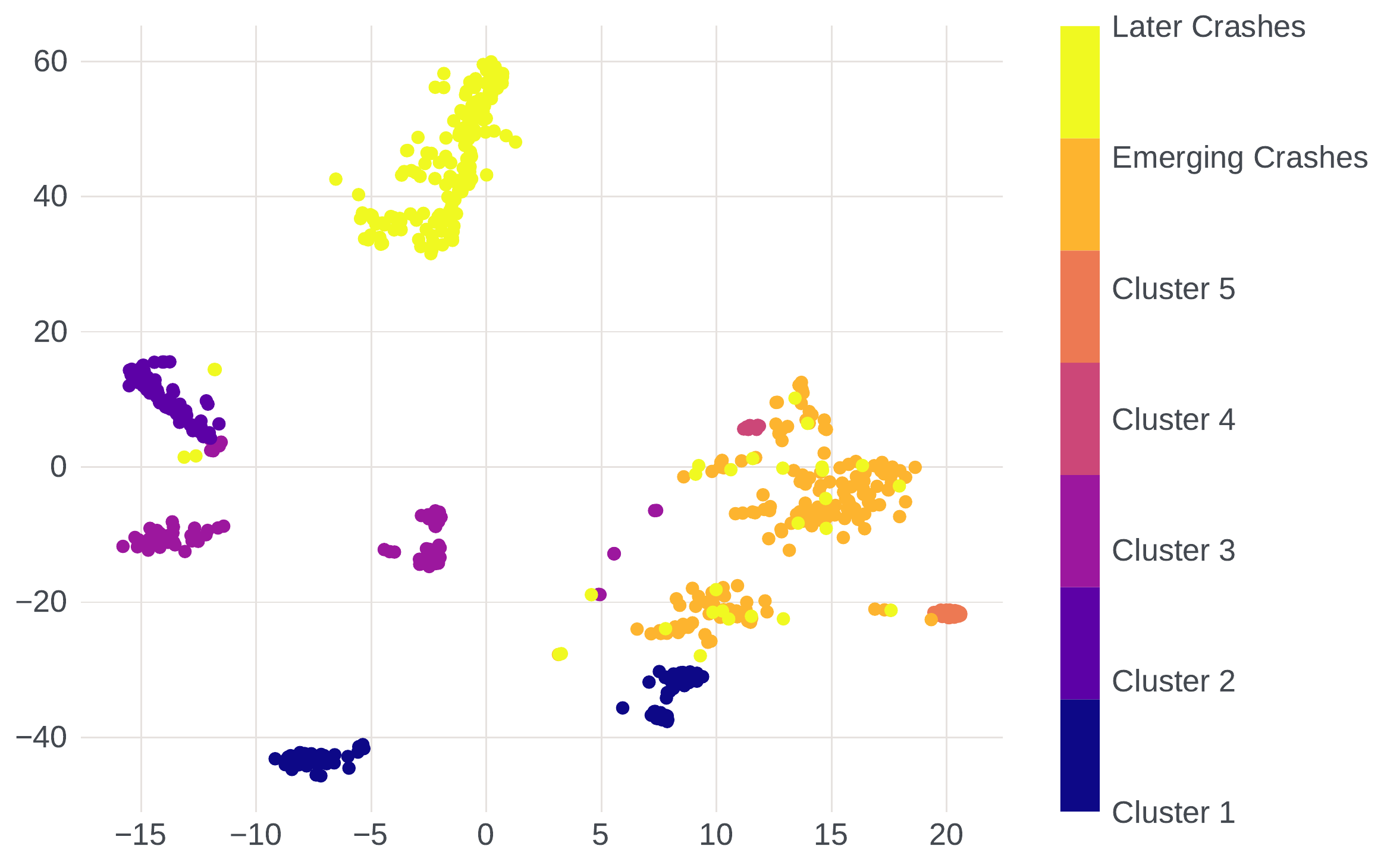}
}

\caption{\kaboom{} crash clustering for version 3xx of the Facebook \ios{} app. We observe that \kaboom{} can identify crashes its has not seen before by clustering them in different parts of the embedding space.}
\label{fig:RQ2}

\end{figure*}

With this research question, we evaluate \kaboom{}'s ability to 
identify and  flag new crash types. 
\Cref{fig:RQ2} presents the results of applying an already trained VAE+DEC model (a)
to an application regression in the middle of the version's lifecycle that has been identified and fixed by engineers (b) 
and a set of crashes late in the examined version's lifecycle.
The initial model was trained with data from the first week of the application's lifecycle.
Note that the K-Means was not re-run with the new regression-related data, so they are not part of a cluster in the diagram.

The \kaboom{} embedding model is able to produce embeddings that set the regression data apart.
The distance of the regression sessions is high in both the \tsne{} space 
and in the original embedding space (not shown in the figure). 
Debug engineers determined that the root cause for this regression 
was related to a particular application view that was allocating memory incorrectly.
\kaboom{} would have been able to flag this emerging regression simply by examining
whether the distance of the embedding vector to existing cluster
centroids is over a pre-determined threshold.

\Cref{fig:RQ2}(c) presents a \tsne{} visualization of the original validation clusters,
the aforementioned regression, and additional crashes collected over a month after
the regression (but still in the same application version). 
The new data from this time period also lies far from both the validation data and the regression-related data. 
Intuitively, it's extremely likely (and indeed happened in this case) that various
memory-related bugs have been fixed and thus our samples of new crashes should shift in
distribution.
From a practical perspective, this means that \kaboom{} should be periodically retrained 
on incoming data, especially since newer crashes are virtually indistinguishable from 
regression-related crashes.

\subsection{RQ3: Explaining cluster membership}

\label{explainability}
\begin{table}
\begin{tabular}{||c || c c || c c c||}
\hline
 Cluster & Rank & Object Name & Presence \%   & Other \% & F1   \\ 
 \hline
 1 & 1 & Obj A & 0.50 & 0.12 & 0.62 \\ 
  & 2 & Obj B & 0.38 & 0.17 & 0.58 \\
  & 3 & Obj C & 0.42 & 0.30 & 0.53  \\
 2 & 1 &  Obj D & 0.72 & 0.08 & 0.64 \\
  & 2 & Obj E & 0.41 & 0.17 & 0.59 \\ 
  & 3 & Obj F & 0.42 & 0.53 & 0.53 \\ 
 3 & 1 & Obj G & 0.12 & 0.01 & 0.64 \\  
  & 2 & Obj H & 0.35 & 0.25 & 0.54\\ 
  & 3 & Obj I & 0.12 & 0.13 & 0.49\\  
 \hline
\end{tabular}
\caption{Example output of presence-based cluster comparison of objects. For each cluster, the top 3 objects are ranked by their F1 score.}
\label{tab:explain}
\end{table}

To explain cluster assignments and provide actionable feedback, 
we add a post processing step of comparing session contents between sessions in clusters. 
The most direct way of assessing object importance is comparing the frequency at which an object is present in the sessions with respect to clusters. 
We define an object to be present in a session if it has any non-zero readings throughout the session. We aggregate the results and rank the top several results per cluster. 
\Cref{tab:explain} is a depiction of how the results look when presented to the engineer.

In \Cref{tab:explain}, for each object, given a cluster, we calculate the percentage of sessions in the cluster with the object present (presence \%). For the same object, we also calculate the percentage of sessions for which it is present across all other clusters combined (other \%). We then calculate an F1-score, given by $F = \frac{presence \%}{presence \% + \frac{1}{2}(presence \% + other \%)}$, which we use to rank most significant objects per cluster in descending order.

In practice, we found that presence-based cluster comparison was a pretty reasonable way of characterizing the important time series in a cluster, for two reasons. The first is that the top-ranked results per group are often related to each other in a clear way, often because they are allocated from the same code files/portions of the application. The second is that, when applied to the emerging crash data, as compared to the original five clusters, the top several objects revealed are all related to the root cause of the emerging crashes.

Likewise, we run the same analysis for the lack of presence. We define an object to be not present in a session if it has all 0 readings for the session. The results for lack of presence do not look as conclusive as the straightforward presence comparisons, mainly due to the sparsity of the OOM data - only a handful of objects are present in all $\approx200$ sessions in the validation set, and a significant amount of objects are present in only one session. Thus, compared to presence comparison, lack of presence comparisons suffer from many more objects to consider and much more noise.

We run analysis on the average values of objects when they are present across the sessions as well. We filter the objects by a minimum threshold of presence first, 5\%, so that objects present in less than 5\% of a given cluster do not skew average results. Then, for each cluster, we compare the average value of of each object against its average value in other clusters. We calculate the mean value of an object for a cluster as the mean value of an object over the timesteps it is present across the sessions it is present in that cluster. Many of the results of average value are different from presence-based analysis, but are not extremely meaningful. The top couple of results contain common objects that are shared across all sessions across clusters, and upon manual inspection didn't relate very closely to any previous OOM issues.

Finally, we also run "mutation tests" on the inputs. For every object in each cluster, we change its values first. If all values of an object are zero, we set the timeseries values to be the average value of the object across all 200 traces (where average is calculated as described in the average value analysis above). If instead the object is present, then we set all values to 0 for that object in the session. We compare the model's clustering of this mutated session against the model's clustering of the non-mutated session, and if it changes, we consider the mutated object important to cluster assignment. In practice, many of the objects that are important to certain clusters overlap with those deemed important by the presence-based analysis. Very rarely does a mutation to average values cause the model to change clusters, but very often the zeroing of objects leads to a change in cluster. Zeroing out objects intuitively should be related to object presence, since the act of zeroing out effectively changes an object from being present to not being present. 

In practice, the results from lack of presence comparison and average value comparisons across clusters do not tend to produce helpful results. Out of the presence comparisons and mutation tests, results from mutation tests seem to be slightly less direct than presence-based comparison, so we prefer the latter, and only display the latter to our engineers.

\section{Discussion and Future Work}

In this section, we revisit our design decisions and 
propose paths to interesting future work.

\paragraph{What is \kaboom{} good for?}
While the \kaboom{} model proved useful in the \oom{} use case,
its embedding model can be useful for several other use cases,
including the following:

\begin{itemize}
    
    \item Cluster participation counting to prioritize handling: when in production,
    \kaboom{} can help engineers prioritize their debugging efforts by simply counting
    how many crashes belong to each cluster.
    
    \item Check for crash types between versions: Engineers can use multiple versions
    of the \kaboom{} model to check whether bugs where fixed in newer versions or
    check whether known crash types where re-introduced.

    \item Real-time identification of emerging crashes: 
    when a minor application version is deployed to clients, 
    \kaboom{} can immediately flag emerging regressions within a few hours.
    \kaboom{} can thus act as an effective alarm to quickly rollback a new release
    if several new application crashes occur.
    
    \item Non-\oom{} crashes: \kaboom{}'s current focus has been \oom{} crashes,
    due to an existing use case. However, there is nothing that prevents \kaboom{}
    to be trained against other types of regressions, for example performance-related
    ones.

\end{itemize}

\paragraph{Input trace sizes} To tackle incoming data sizes, 
\kaboom{} is currently restricted to processing the last 100 timesteps before a crash.
The actual root cause for a crash however may not manifest in the last 100 timesteps.
For example, a misbehaving class might have allocated 1000s of objects in the past
resulting in increasing memory pressure.
While the class timeseries will be part of the processed trace, 
the model may be learning patterns relating to increasing values to other timeseries.
\kaboom{} must come up with a compact 1-dimensional representation 
of large, sparse 2-dimensional matrices,
while also exploiting the temporal properties of the input.
To solve this problem, we have experimented with sliding a window of size $n$,
along the input time $t$ axis (producing $t - n $ inputs), and training a recurrent
autoencoder to reconstruct the last timestep of each window, to no avail.
The recently introduced Interfusion model~\cite{li2021multivariate} 
attempts to solve this problem by coming up with two embeddings, 
one per the timing and metrics axis, respectively,
using a sliding window over the input. 
Perhaps combining those two embeddings using a further recurrent layer may
prove useful for \kaboom{}.

\paragraph{Model architecture}
There are various other architectural designs that may improve \kaboom{} performance. 
On the top of our list lay real and fake discrimination, and attention mechanisms.

The first extension is an addition of a discrimination task while training the encoder portion of the network, as in references~\cite{ranjan2018fake, ma2019learning}. 
The discrimination task used by these approaches is similar to 
the discriminator module employed by generative adversarial networks.
The task would introduce "fake" examples generated by sampling non-crashing sessions,
which then are encoded through the encoder and a separate decoder network then predicts 
whether that input was a "real" sample from the dataset or a "fake" one. 
This approach could also be seen as a way to further regularize the hidden 
representation via multi-task learning ~\cite{evgeniou2004regularized}.

Though attention mechanisms rose to prominence 
for sequence-to-sequence tasks~\cite{vaswani2017attention},  
different time series models have also began incorporating attention mechanisms 
for forecasting ~\cite{shih2019temporal, liang2018geoman, huang2019dsanet} and 
classification ~\cite{song2018attend, karim2017lstm}. 
Beyond potentially improving the quality of representations learned, the addition of attention could provide more interpretability directly into the decisions 
the model is making, aiding the visual results that \kaboom{} presents to the debugging
engineers. 

\paragraph{What is a good clustering?} 
\kaboom{} optimizes clustering for incoming data distributions,
but how can we know that the resulting clustering is good for engineers?
Before involving end users, one way to explore this is by checking
intra-cluster consistency by analyzing the most popular objects within all
sessions in a cluster. 
Our experience shows that presence-based cluster comparison can help both engineers and
researchers working on the clustering model to gain confidence in the model's outputs.

\paragraph{Efficient deployment} 
\kaboom{} models need to be re-trained and deployed periodically, 
as a result of object value drift as bugs are resolved in new application versions (discussed in \ref{RQ2}). 
This results in overhead in training and maintaining models. More investigation is needed on the nature of how metrics drift over time as bugs are fixed and new code is pushed. Depending on exactly how object time series drift over time, further research could focus on the feasibility of online model training as crash data is collected, which would eliminate the need for multiple models and save manual effort in retraining models.


\section{Related Work}

There has been relatively little work done in clustering multivariate timeseries to perform incident analysis. From a model point-of-view, there exists similar work that leverage neural networks to generate embeddings to detect anomalies and catch incidents before they happen or address them after they happen. From a problem point-of-view, there have been several works that employ time-series clustering to software incident analysis, but rely on traditional notions of clustering.

Recent research in incident anomaly detection has incorporated recurrent neural networks ~\cite{islam2020anomaly} and convolutional neural networks ~\cite{munir2018deepant} to have superior performance on problems with plentiful data and little prior knowledge compared to traditional methods of time series clustering. Additionally, modifications to these types of networks allow for addition of stochastic units in the model to capture some of the randomness in input noise, with a common application being in variational auto-encoders. The most recent algorithms, such as OmniAnomaly ~\cite{su2019robust} and Interfusion ~\cite{li2021multivariate}, combine varational auto-encoder models with recurrent networks and convolutional networks to achieve state of the art performance across a plethora of datasets, including a server machine dataset ~\cite{su2019robust}. While several of these models also use neural networks to generate embeddings, because the embeddings are trained to optimize for the anomaly detection problem, simply appending a clustering step to those architectures is not enough to produce \kaboom{}-like results. In particular, the representations that several of these models learn try to collapse the learned representations into a single, dense space, such that they would only produce one cluster across the whole dataset.

For the relatively few works where timeseries clustering is used for incident analysis, clustering appears primarily as a prior step to anomaly detection. 
Both Li et. al. ~\cite{li2018robust} and Qian et. al. ~\cite{qian2020large}
cluster key performance indicator (KPI) time series to figure out common patterns, 
as input to their anomaly detection algorithms. 
Both papers explore correlation-based methods to compute distance between KPI time series,
and then apply traditional clustering algorithms such as DBSCAN. 
We are not aware of any further research that directly applies clustering
for service monitoring or regression detection to directly learn and reason about the underlying data. Thus, \kaboom{} is the first work to combine models from the traditional timeseries clustering space, and apply it to a service monitoring setting with multivariate data.

\section{Conclusion}
In this paper, we present a new technique, timeseries fingerprinting, 
for tackling the problem of unsupervised crash categorization at scale.
\kaboom{} trains an autoencoder model to learn to embed incoming crashes,
represented by multivariate timeseries of object allocations, 
into a much smaller dimensional space, where traditional distances between points are easy to compute. The embeddings are learned in such a way that crashes with related objects and thus behavior lie close together, and unrelated crashes lie far apart. 
Moreover, \kaboom{} learns cluster centroids tuned to the incoming data distributions,
which enables it to both separate different types of crashes 
and to identify when new types of crashes occur.
Finally, \kaboom{} offers intuitive explanations
for what types of patterns are common to groups of crashes, 
based on contrastive examination of the objects that comprise them.
Based on our current experiences and developer feedback,
in future iterations, we plan 
to train \kaboom{} to compress arbitrarily sized input traces,
to enhance the cluster explanations using causal learning, and
to apply \kaboom{} for post-mortem analysis of services.

\bibliographystyle{ACM-Reference-Format}
\bibliography{kaboom}

\end{document}